\documentclass[sn-aps]{sn-jnl}
\usepackage{graphicx}%
\usepackage{multirow}%
\usepackage{amsmath,amssymb,amsfonts,wasysym}%
\usepackage{braket,amsfonts}%
\usepackage{dsfont}%
\usepackage{amsthm}%
\usepackage{mathrsfs}%
\usepackage[title]{appendix}%
\usepackage{xcolor}%
\usepackage{textcomp}%
\usepackage{manyfoot}%
\usepackage{booktabs}%
\usepackage{algorithm}%
\usepackage{algorithmicx}%
\usepackage{algpseudocode}%
\usepackage{listings}%
\usepackage{hyperref}

\allowdisplaybreaks
\newcommand{\be}{\begin{equation}}
\newcommand{\ee}{\end{equation}}
\newcommand{\bga}{\begin{gather}}
\newcommand{\ega}{\end{gather}}
\newcommand{\bea}{\begin{eqnarray}}
\newcommand{\eea}{\end{eqnarray}}
\newcommand{\dagga}{{\phantom{\dagger}}}
\newcommand{\bR}{\mathbf{R}}

\newcommand{\bk}{\mathbf{k}}

\newcommand{\bGamma}{\mathbf{\Gamma}}
\newcommand{\bX}{\mathbf{X}}

\newcommand{\bM}{\mathbf{M}}
\newcommand{\Ima}{{\text{Im}}}
\newcommand{\Rea}{{\text{Re}}}
\newcommand{\dis}{\displaystyle}

\newcommand{\up}{\uparrow}
\newcommand{\down}{\downarrow}
\newcommand{\fract}[2]{\frac{\dis \;#1\;}{\dis \;#2\;}}

\newcommand{\eqn}[1]{(\ref{#1})}

\newcommand{\ep}{{\epsilon}}

\newcommand{\bw}{\begin{widetext}}
\newcommand{\ew}{\end{widetext}}
\newenvironment{eqs}%
{\begin{equation} \begin{aligned}}%
{\end{aligned} \end{equation} }
\newcommand{\beal}{\begin{eqs}}
\newcommand{\eal}{\end{eqs}}
\newcommand{\bd}[1]{{\boldsymbol{#1}}}

\begin{document}
\title{
Luttinger surface dominance and Fermi liquid behaviour of topological Kondo insulators  SmB$_6$ and YbB$_{12}$}

\author[1]{Andrea Blason}\email{ablason@sissa.it}
\author[1]{Ivan Pasqua}\email{ipasqua@sissa.it}
\author[2,3]{Michel Ferrero}\email{michel.ferrero@polytechnique.edu}
\author[1]{Michele Fabrizio}\email{fabrizio@sissa.it}

\affil[1]{International School for Advanced Studies (SISSA), Via Bonomea 265, I-34136 Trieste, Italy} 

\affil[2]{CPHT, CNRS, Ecole Polytechnique, Institut Polytechnique de Paris, 91128 Palaiseau, France}
\affil[3]{Coll\'ege de France, 11 place Marcelin Berthelot, 75005 Paris, France}

\abstract{\textbf{Defying the traditional classification into metals and insulators, several materials  
simultaneously display metallic thermal properties  
and insulating electric behaviour, as if they hosted quasiparticles carrying entropy but not charge. Among them, some materials 
also possess quantum oscillations in magnetic fields as if they had well-defined Fermi surfaces despite the insulating gap. This remarkable dichotomy has
been observed in the topological Kondo insulators SmB$_6$ and YbB$_{12}$. 
Prompted by the peculiar mixed-valence nature of these compounds, involving $f$ and $d$ electrons of the lanthanide, we propose an  
explanation of their intriguing properties drawing inspiration from the physics of the pseudogap phase in underdoped cuprates. We argue that the $f$ and $d$ subsystems, when considered separately, act, respectively, as electron- and hole-doped Mott insulators, featuring Fermi pockets coexisting with Luttinger surfaces responsible for the pseudogap.
When the two are coupled to each other a hybridisation gap opens up, and the whole turns into a topological insulator endowed with genuine chiral edge states. 
However, the Luttinger surfaces persist and  
support neutral quasiparticles.
This scenario, supported by numerical simulations within the dynamical cluster approximation, effectively 
resolves the paradoxical phenomenology of SmB$_6$ and YbB$_{12}$.
}}

\maketitle

\noindent
The low energy physics of the topological Kondo insulators (TKIs) SmB$_6$ and YbB$_{12}$ is characterised by opposite-parity localised  $4f$ and itinerant $5d$ rare earth orbitals mutually coupled by an inter-site hybridisation, thus possessing the prerequisites to realise $Z_2$ topological insulators~\cite{Coleman2010, Coleman_review}.\\
These materials display a low-temperature size-independent resistivity plateau hinting at the presence of conducting surface states \cite{NANBA1993440, PhysRevB.86.075105, Kim2013, PhysRevX.3.041024, Nakajima2016,Hlawenka2018,Ohtsubo2019,Ohtsubo2022,Hagiwara2016, Sato_2021, PhysRevB.107.165132} that are generally interpreted as chiral edge modes arising from a non-trivial bulk topology, although a decisive experimental confirmation is still lacking. \\
Narrow-gap insulators in DC transport, the optical conductivity in 
YbB$_{12}$ \cite{Okamura-PRB1998} and especially in SmB$_6$ \cite{Armitage-PRB2016} even shows an abundance 
of dipole-active in-gap excitations.  
Contrary to 
expectations, SmB$_6$ and YbB$_{12}$ 
nonetheless exhibit metal-like bulk thermal properties. SmB$_6$ shows 
temperature-linear specific heat~\cite{PhysRevB.96.115101, PhysRevX.4.031012, FLACHBART2006610} and thermal conductivity~\cite{Hartstein2018},  
(the latter  
still debated~\cite{PhysRevLett.116.246403, PhysRevB.97.245141}).  
YbB$_{12}$ displays unambiguous linearity in both quantities~\cite{Sato2019}.

Notoriously, SmB$_6$ displays unexpected quantum oscillations in magnetisation \cite{science_qo, Hartstein2018, Hartstein2020, LiScience} and in specific heat~\cite{LaBarre_2022}, but not in resistivity, whereas YbB$_{12}$ shows oscillations in both magnetisation \cite{Liu_2018, Xiang2021} and resistivity \cite{resistivityQO, PhysRevX.12.021050, Xiang2021}.
The enigmatic physical scenario arising from these TKIs can thus be articulated as a clear puzzle: how can we reconcile an insulating state, potentially topological, with these other Fermi liquid features typical of metals? 
\\

\noindent
To address this striking inconsistency, several theoretical explanations have been proposed, including the role of magneto-excitons~\cite{Cooper}, the possibility of quantum oscillations in narrow gap insulators~\cite{CooperQO, ZhangQO, PandaQO}, the enhancement of quantum oscillations in Kondo insulators~\cite{LuQO}, or the interplay between surface states and the Kondo breakdown~\cite{ColemanQO}. However, these proposals predict substantial deviations from the Lifshitz-Kosevich theory of quantum oscillations \cite{Lifshitz} that have not been observed. In addition, a crucial role of vacancies and impurities has also been advocated \cite{Pagliuso-PRR2020,Armitage-PhysicaB2018,Paglione-Science2023}, though it cannot explain the metallic heat transport.  
Alternative, less conventional explanations involve the existence of Majorana-like excitations 
\cite{Baskaran-2015,Coleman-PRL2017,Varma-2023}, whose measurable signatures have until now eluded experimental observations. 
A different proposal is that of a neutral Fermi surface originating from exotic composite excitons~\cite{PhysRevB.97.045152,Chowdhury2018}, which faces issues in explaining the conducting surface. In addition, this proposal predicts deviations of the thermal conductivity from a linear-in-temperature behaviour that have not been observed. 
Therefore, a comprehensive interpretation of the ambiguous properties of SmB$_6$ and YbB$_{12}$, partly insulating and partly metallic, is still lacking. This is precisely the goal of the present paper.\\

\begin{figure*}
\centerline{\includegraphics[width=1.0\textwidth]{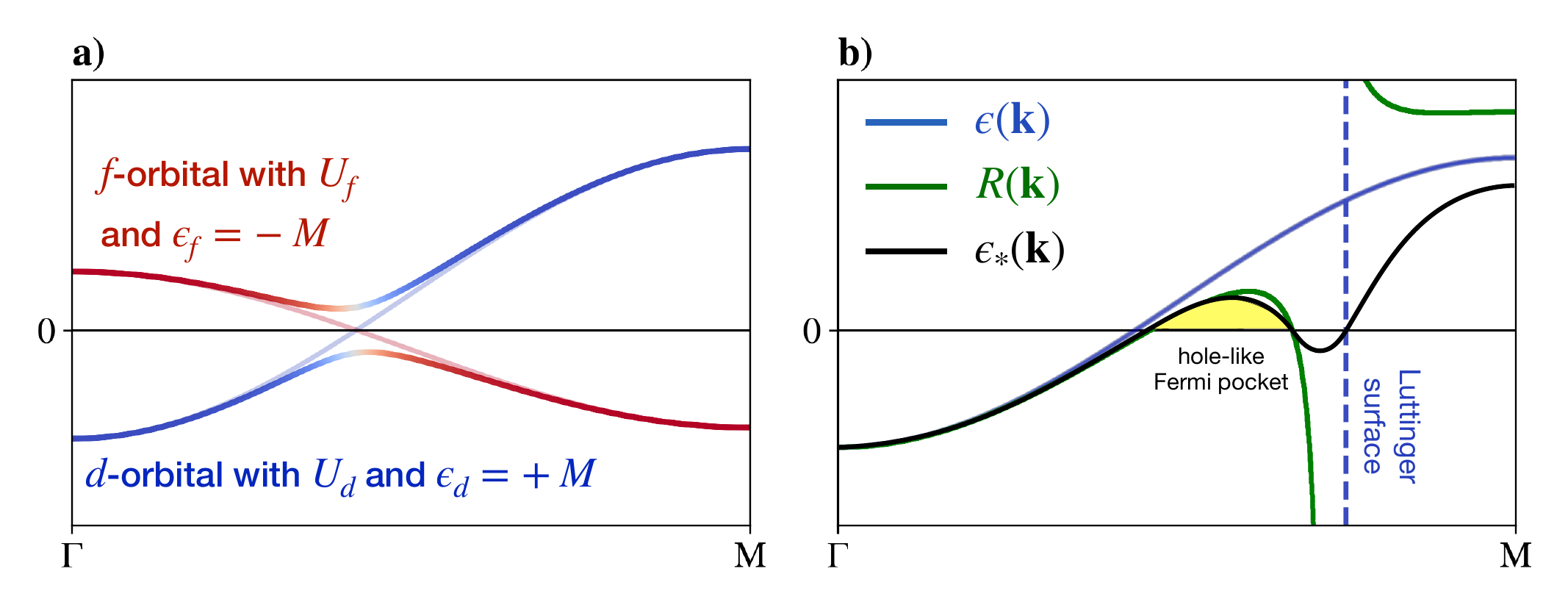}}
\caption{Panel (a): Band structure of a model weakly-correlated  
topological insulator along some 
high-symmetry direction. The two opposite-parity orbitals feel their own Hubbard repulsion, $U_f$ and $U_d$, and are 
split in energy by $2M$. The crossing of the two bands becomes an avoided one because of the inter orbital hybridisation. Panel (b): Effective energy $R(\bk) = \ep(\bk)-\mu + \Sigma(0,\bk)$, in green, and, in black, the quasiparticle one, $\ep_*(\bk)$ in \eqn{quasiparticle energy single band}, of a single-band model in presence of a Luttinger surface, i.e., of a simple pole of $\Sigma(0,\bk)$. The blue line is the non-interacting band dispersion, and the horizontal solid line is the zero of energy that corresponds to the chemical potential.
}
\label{figure1}
\end{figure*}

\noindent
Kondo insulators are commonly discussed in terms of periodic Anderson models \cite{PWA-PR1961} where only the $f$ electrons feel a strong on-site repulsion whereas the conduction bands are assumed non-interacting. However, such description might fail in SmB$_6$ and YbB$_{12}$ where both $4f$ and conduction $5d$ electrons belong to the same lanthanide ion and are therefore expected to be substantially entangled with each other by the Coulomb repulsion \cite{Chowdhury2018,Varma-2023}, see also the Supplementary Material. 
This is consistent with the observed valence fluctuation of the lanthanides, $f^n\,d^0 \leftrightarrow f^{n-1}\,d^1$, $n=6$ and $n=14$ in SmB$_6$ \cite{Mixed-valence-Sm} and YbB$_{12}$ \cite{Mixed-valence-Yb}, respectively, 
with the $f^{n-2}\,d^2$ configuration not seen. Since $f^6$ and $f^{14}$ are both non-magnetic, the minimal model with just single $f$ and $d$ 
orbitals implies the mapping $f^n\,d^0$ onto $f^2\,d^0$, non-magnetic, and $f^{n-1}\,d^1$ onto $f^1\,d^1$, magnetic, whereas 
$f^0\,d^2$ is suppressed by correlations. This is equivalent to assuming two different 
single-band Hubbard models, one based in the odd-parity $f$ orbital,  
the other in the 
even-parity $d$. 
These orbitals are mutually coupled by spin-dependent inter-site hybridisation, charge repulsion and Coulomb exchange. The density must equal two electrons per site, thus occupation numbers $n_f=1+\delta$ and $n_d=1-\delta$, where $0<\delta< 1$ is enforced by different orbital energies, so that the dominant valence fluctuation for strong enough interaction is indeed $f^2\,d^0\leftrightarrow f^1\,d^1$. 
In Fig.~\ref{figure1}(a) we sketch a possible weak-coupling band structure along a  
high-symmetry direction in $\bk$-space. 
Upon increasing the interaction strength, each Hubbard model is driven towards a Mott insulating phase with vanishingly small $\delta$. Since the inter-orbital hybridisation is finite, we expect a single Mott transition with the simultaneous localisation of both orbitals \cite{OSMT_Kotliar}. Therefore, each Hubbard model effectively behaves before the transition as a weakly-doped Mott insulator, hole-doped for the $d$ and electron-doped for the $f$. We emphasise that the above description in terms of  
one $f$ and one $d$ orbitals implicitly assumes that the Mott insulator corresponds to the magnetic 
$f^{n-1} d^1$ configuration of SmB$_6$ and YbB$_{12}$. We postpone a more extensive discussion of this picture 
to the end of the paper, assuming in the following the two-orbital representation. \\

\noindent
Let us at first discuss the low-doping and low-temperature regime in a single-band Mott insulator. 
In the context of underdoped cuprates, it has been proposed \cite{Rice-PRB2006,Rice-RPP2011,Alexei-RPP2019} and observed in several model calculations
\cite{PhysRevB.74.125110,cluster1,cluster2,PhysRevLett.102.056404,PhysRevB.82.134505,Sordi-PRL2010,PhysRevB.82.155101,PhysRevB.83.214522,Federico-PRB2012,PhysRevLett.110.216405,Georges-PNAS2018,Georges-PRX2018,PhysRevLett.120.067002,Imada-NatComm2023} that a weakly-doped Mott insulator is characterised by a pseudogap in the single-particle spectrum. Concurrently, Luttinger's theorem \cite{Luttinger} is 
violated and the Fermi surface just accounts for the doping away from half-filling. 
We take for granted the existence of such pseudogap and further assume that it is due to the 
presence of a Luttinger surface (LS) \cite{Igor-PRB2003}. The LS is the manifold in the Brillouin zone (BZ) of the 
zeros of the Green's function $G(i\ep,\bk)$, or, equivalently, the poles of the self-energy 
$\Sigma(i\ep,\bk)$, at zero Matsubara frequency, $\ep=0$. A LS is indeed known to yield \textcolor{blue}{a} violation of Luttinger's theorem~\cite{Heath_2020}: in a single-band model it accommodates one electron per site \cite{Jan-PRB2022} whatever its volume, which implies the existence 
of Fermi pockets that account for the missing electron density.  It has been also shown that a LS can host Landau quasiparticles \cite{Michele2}, which are invisible in the single-particle spectrum. Thus, while they do not contribute to electric transport, they nonetheless yield Fermi-liquid-like thermal and magnetic properties, including quantum oscillations 
in magnetic field \cite{Michele3}, see also the Supplementary Material. These quasiparticles legitimately exist 
in Mott insulators, where they are neutral excitations, 
Anderson's \textit{spinons} \cite{PWA-RVB,PWA-PRL1987}, whose "Fermi surface" 
is therefore the Luttinger surface. Specifically, the thermal Green's function of a single-band model of interacting electrons satisfies Dyson's equation
\beal
G(i\ep,\bk)^{-1} = i\ep +\mu - \ep(\bk) -\Sigma(i\ep,\bk)\,,
\label{Dyson equation}
\eal
where $\mu$ is the chemical potential, $\ep(\bk)$ the non-interacting dispersion and 
$\Sigma(i\ep,\bk)= \Sigma(-i\ep,\bk)^*$ the self-energy. Following \cite{Michele3}, one defines 
\beal
Z(\ep,\bk)^{-1} = 1 - \fract{\;\Ima\,\Sigma(i\ep,\bk)\;}{\ep}\,,
\label{quasiparticle residue single band}
\eal
where $Z(\ep,\bk)=Z(-\ep,\bk)\in [0,1]$ and $Z(\bk)=Z(0,\bk)$ is the quasiparticle residue. Correspondingly, the quasiparticle energy is 
\beal
\ep_*(\bk) &= Z(\bk)\,\Big(\ep(\bk)-\mu + \Sigma(0,\bk)\Big)\equiv Z(\bk) \,  R(\bk) \, \in \mathbb{R} \, .
\label{quasiparticle energy single band}
\eal 
In this formalism, the Luttinger surface corresponds to the roots of $Z(\bk)$, whereas the Fermi surface 
to those of the effective energy $R(\bk)$, i.e., the poles of $G(0,\bk)$, see \eqn{Dyson equation}. We note that if $\Sigma(0,\bk)$ has a simple pole on the LS, then $Z(\bk)$ has a second-order zero. It follows that the quasiparticle energy $\ep_*(\bk)$ vanishes on the Luttinger surface as well as on the Fermi surface, both of which therefore define the \textit{quasiparticle Fermi surface}. In Fig.~\ref{figure1}(b) we sketch the behaviour of $R(\bk)$ and $\ep_*(\bk)$ when a LS is present. The physics of the pseudogap, thus of the singular self-energy and of Luttinger's theorem breakdown, is presumably that of a failed symmetry breaking state, likely due to strong short-range antiferromagnetic correlations \cite{Altshuler-EPL1998} in weakly-doped Mott insulators. Moreover, it also provides a simple mechanism of a continuous non-symmetry breaking Mott transition: the Fermi pockets gradual disappear, leaving the LS untouched. We mention that a similar scenario has been recently advocated 
in the context of symmetric mass generation \cite{PhysRevB.107.195133,PhysRevB.108.205117,PhysRevLett.132.081903,su2024global}.\\ 

\noindent
Let us extend the above picture to the case of  
interest, i.e., two coupled Hubbard models. In absence of 
inter-orbital hybridisation, each model describes a weakly-doped Mott insulator with its own Luttinger surface and Fermi pockets. Specifically, one model must have electron-like Fermi pockets and the other hole-like 
ones, with equal enclosed volumes. Turning on the inter-orbital hybridisation, one can imagine that the Fermi pockets will get gapped, while the Luttinger surfaces survive. 
Since the latter do not contribute to electric transport, the 
system describes a charge insulator which, however, may display thermal properties of a metal as well as quantum oscillations \cite{Michele3}, which is 
precisely the phenomenology observed in SmB$_6$ and YbB$_{12}$. \\

\noindent
To further explore this scenario and confirm its potential feasibility, we first study the two-band 
Hubbard model assuming $d$ and $f$ orbitals  
to be related to each other by a particle-hole transformation, an artifice which is numerically very convenient, see below. 
The corresponding single-particle Hamiltonian reads \cite{PhysRevLett.114.177202,PhysRevB.85.125128,Assad1}
\beal
H_0(\bk) &= \big(M+\epsilon(\bk)\big) \, \tau_3 +
 2 \, \lambda \, \tau_1 \, \big(\, \sin k_x \, \sigma_1 + \sin k_y  \, \sigma_2 \, \big)\, ,
 \label{H0_ph}
\eal
where the Pauli matrices $\sigma$'s and $\tau$'s act, respectively, in the (pseudo) spin, which represents specific components of the total angular moment, and orbital subspaces, with $d$ and $f$ corresponding to $\tau_3=+1$ and $\tau_3=-1$, respectively, M is a constant, and $\ep(\bk)=-2\,t\,(\cos k_x + \cos k_y)$. We take $t=1$ as the energy unit and $\lambda = 0.1$.
The Hamiltonian \eqref{H0_ph} has particle-hole symmetry and preserves parity, crucial for defining the $Z_2$ topological invariant. At the four high-symmetry points, $\mathbf{\Gamma}=(0,0)$, $\mathbf{X}=(\pi,0)$, $\mathbf{Y}=(0,\pi)$, and $\mathbf{M}=(\pi,\pi)$, the inter-orbital hybridisation vanishes, leading to Bloch waves with well-defined parity. The system topology is hence governed by the band inversions within the BZ. Notably, $H_0(\mathbf{k})$ admits three distinct topological phases as a function of $M$. For $|M| > 4$, the model is a trivial insulator, otherwise it exhibits two non-trivial topological phases with degenerate bands that carry opposite Chern number depending on the sign of $M$.
Next, we introduce the simplest intra-orbital Hubbard repulsion which 
we write as 
\be
H_U = \fract{U}{2}\, \sum_\bR  \,\sum_{a=d,f}\,\Big(n_{a\bR\uparrow}+n_{a\bR\downarrow}-1\Big)^2  \, ,
\label{Hint_ph}
\ee
to make particle-hole symmetry explicit,  
where $n_{f\bR\sigma} =f^\dagger_{\bR,\sigma}\, f^\dagga_{\bR,\sigma}$ and   
$n_{d\bR\sigma} =d^\dagger_{\bR,\sigma}\, d^\dagga_{\bR,\sigma}$. \\
In this two-band case the Green's function 
and the self-energy become $4\times 4$ matrices, 
$G(i\ep,\bk)= G(-i\ep,\bk)^\dagger$ 
and $\Sigma(i\ep,\bk)= \Sigma(-i\ep,\bk)^\dagger$ . Equation \eqn{quasiparticle residue single band} now reads 
\beal
Z(\ep,\bk) &= \Bigg(
1 - \fract{\;\Sigma(i\ep,\bk)-\Sigma(i\ep,\bk)^\dagger\;}{2i\ep}\Bigg)^{-1}
= A(\ep,\bk)^\dagger\,A(\ep,\bk)\,,\label{Z-matrix}
\eal
with $Z(\ep,\bk)$ semi positive definite, thus the last equation, 
with eigenvalues in the interval $[0,1]$, while \eqn{quasiparticle energy single band} turns into the 
quasiparticle Hamiltonian matrix \cite{Andrea-PRB2023}
\beal \label{Hqp}
H_*(\bk) = H_*(\bk)^\dagger &= A(\ep,\bk)\,
\Big( H_0(\bk) -\mu + \Sigma(0,\bk)\Big)\,A(\ep,\bk)^\dagger \\
&\equiv A(\ep,\bk) \; R(\bk) \; A(\ep,\bk)^\dagger \, ,
\eal 
with $R(\bk)$ the renormalised Hamiltonian matrix.\\
We solve the interacting model, \eqn{H0_ph} plus \eqn{Hint_ph}, using the dynamical cluster approximation (DCA) \cite{RevModPhys.77.1027,SETH2016274,PARCOLLET2015398}, a cluster extension of dynamical mean field theory that yields a non-local self-energy (see Methods section). Exploiting the particle-hole symmetry, we can afford a DCA calculation with four patches centred at the high-symmetry points $\mathbf{\Gamma}$, $\mathbf{X}$, $\mathbf{Y}$ and $\mathbf{M}$ \cite{PhysRevB.82.155101}, thus a good resolution in momentum space. The reason is that at those points the Green's function is diagonal by symmetry, allowing us to treat just a single orbital, the other being obtained by a particle-hole transformation \footnote{This simplified four-patch approach is extremely convenient even if it does not give direct access to the off-diagonal elements of $\Sigma(i\ep,\bk)$  
away from the high-symmetry points. That would require at least eight patches, beyond current capabilities. Whenever necessary, we shall discuss  
the possible consequences of those off-diagonal elements. 
Another issue is that DCA yields a self-energy that is a step function in $\bk$. 
There are ways to interpolate such piecewise-constant function so to obtain an approximate self-energy smooth in momentum space. Here, we adopt a scheme that is similar to the cumulant interpolation \cite{PhysRevB.74.125110,cluster2}.}.\\

\begin{figure*}
\centering
\includegraphics[width=1.0\textwidth]{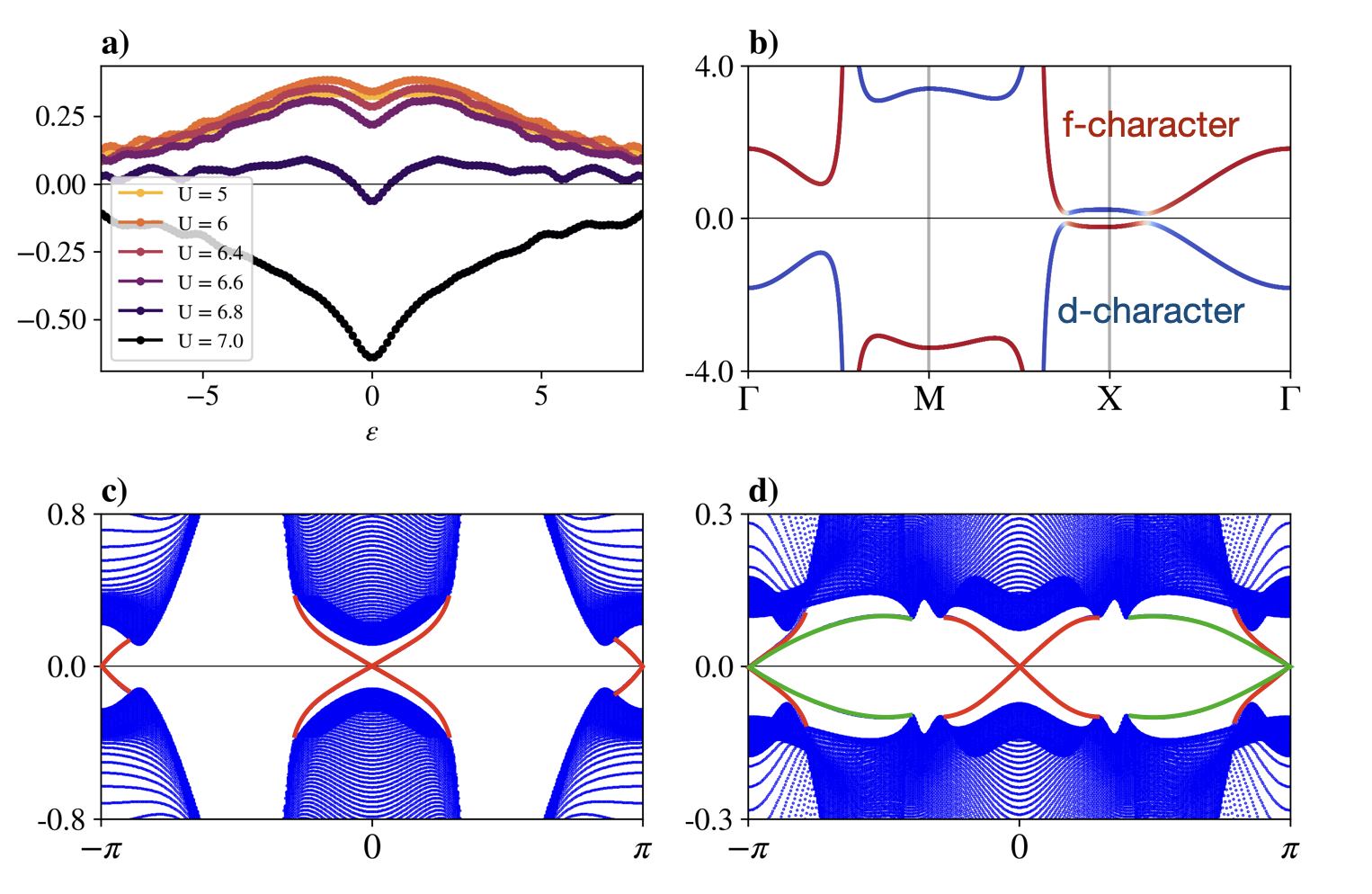}
\caption{ Panel (a):  $\Rea\Sigma_{ff}(i \ep, \mathbf{M} ) - \Sigma^{HF}_{ff}$ as function of the Matsubara frequency and for different values of $U$. For $U = 5$, this quantity is smooth and monotonously decreases in $|\ep|$. At $U \sim 6$ it develops a minimum at $\ep=0$ that grows with $U$. 
Panel (b): Eigenvalues of the renormalised Hamiltonian $R(\bk)$ in Eq.~\eqn{Hqp} obtained from the 
interpolation scheme discussed in the text at $U=6.8$. 
The colours of the lines represent the orbital character of the eigenvectors. Two avoided crossing between the Fermi pockets along $\bGamma \to \bX \to \bM$ are present, along with a divergence along $\bGamma \to \bM \to \bX$ due to the Luttinger surface. 
Panel (c): Eigenvalues of $R(\bk)$ with OBC along \textit{y} adding a tiny off-diagonal element in 
$H_1(\bk)$ of Eq.~\eqn{self-energy-ansatz}. We note the two pairs of chiral edge states in red. 
Panel (d): Eigenvalues of the quasiparticle Hamiltonian $H_{*}(\mathbf{k})$ for the same self-energy as in panel (b). In this case, along with the presence of genuine edge states, still in red, which are surface poles of the Green's function, there are also new edge states, in green, which are surface zeros of the Green's function and originate from the gapping of the Luttinger surfaces. }
\label{figure2}
\end{figure*}

In Fig.~\ref{figure2}(a) we plot 
$\Delta\Sigma_{ff}(i\ep) \equiv \Rea\Sigma_{ff}(i\ep,\bM) - \Sigma_{ff}^{HF}$ 
in the patch $\bM$ as function of $\ep$, at $M=1.8$, thus within the uncorrelated topological phase, $\beta=50$ and for different values of $U$, where $\Sigma_{ff}^{HF}$ is the Hartree-Fock contribution, i.e., the $|\ep|\to\infty$ limit of the self-energy. We observe that 
$\Delta\Sigma_{ff}(i\ep) = -\Delta\Sigma_{dd}(i\ep)$ looks like the difference of a very broad Lorentzian-like function minus a narrower one that appears above $U=5$ and grows with further increase of $U$. 
This behaviour suggests that we 
may model the self-energy matrix 
at small Matsubara frequencies and large $U>5$ in the form \cite{Giorgio-NatComm2023}
\beal
\Sigma(i\ep,\bk) &\simeq \fract{\Delta^2}{\;i\ep + H_1(\bk)\;}\; + \tilde{\Sigma}(\bk),
\label{self-energy-ansatz}
\eal      
where the first term describes the narrow peak in Fig.~\ref{figure2}(a), with 
$H_1(\mathbf{k})$ resembling $H_0(\mathbf{k})$ though with renormalised parameters, and 
$\tilde{\Sigma}(\bk)$ the $\ep=0$ value of the broad part. At $U=6.8$, we can safely take 
$\tilde{\Sigma}(\bk)\simeq \Sigma^{HF}$, see Fig.~\ref{figure2}(a), and thus fit the diagonal elements of $H_1(\mathbf{k})$ from the cluster self-energy in the four patches. 
In this way we obtain the eigenvalues of $R(\bk)$ shown in Fig.~\ref{figure2}(b) that display both singular points, corresponding to poles of the self-energy, and visible avoided crossings between the Fermi pockets of the two orbitals centred around $\mathbf{X}$ and $\mathbf{Y}$. We note that the coincident Luttinger surfaces of $d$ and $f$ electrons remain unaltered because the self-energy lacks off-diagonal elements within our four-patches dynamical cluster approximation. Indeed, since the quasiparticle residue at the LS vanishes, the bare inter-orbital hybridisation is ineffective and only off-diagonal terms of the self-energy could split the $f$ and $d$ Luttinger pockets opening a gap. Looking at equations \eqn{Z-matrix} and \eqn{Hqp}, one realises that the gapping of the $f$ and $d$ Luttinger surfaces may occur only when they cross, otherwise the vanishing quasiparticle residue protects their existence. In the present case, where the two surfaces are on top of each other by particle-hole symmetry, the gapping is unavoidable. To reproduce it, we add a tiny off-diagonal element in $H_1(\mathbf{k})$ that provides the expected off-diagonal component of the self-energy. With that 
additional term, we obtain $R(\bk)$ of Fig.~\ref{figure2}(c), calculated in open boundary conditions (OBC), which is a simple hallmark of the bulk topology. We note two pairs of in-gap chiral modes crossing the chemical potential that correspond to edge poles of the Green's function. The quasiparticle energies \eqn{Hqp} in OBC, Fig.~\ref{figure2}(d), shows, besides these edge states, additional surface modes that correspond to edge zeros of the Green's function 
\cite{Gurarie-zeros-PRB2011,Essing&Gurarie-PRB2011,Giorgio-NatComm2023,wagner2023edge} caused by  
gapping of the bulk Luttinger surfaces. We remark that the Chern numbers of the renormalised bands are the same as the non-interacting ones, as also testified by the edge modes, two of them being poles and one a zero of the surface Green's function, which suggests that no gap closing is needed to connect this \textit{topological pseudogap insulator} (TPI) to the non-interacting one. \\

\begin{figure*}
\centerline{\includegraphics[width=1.0\textwidth]{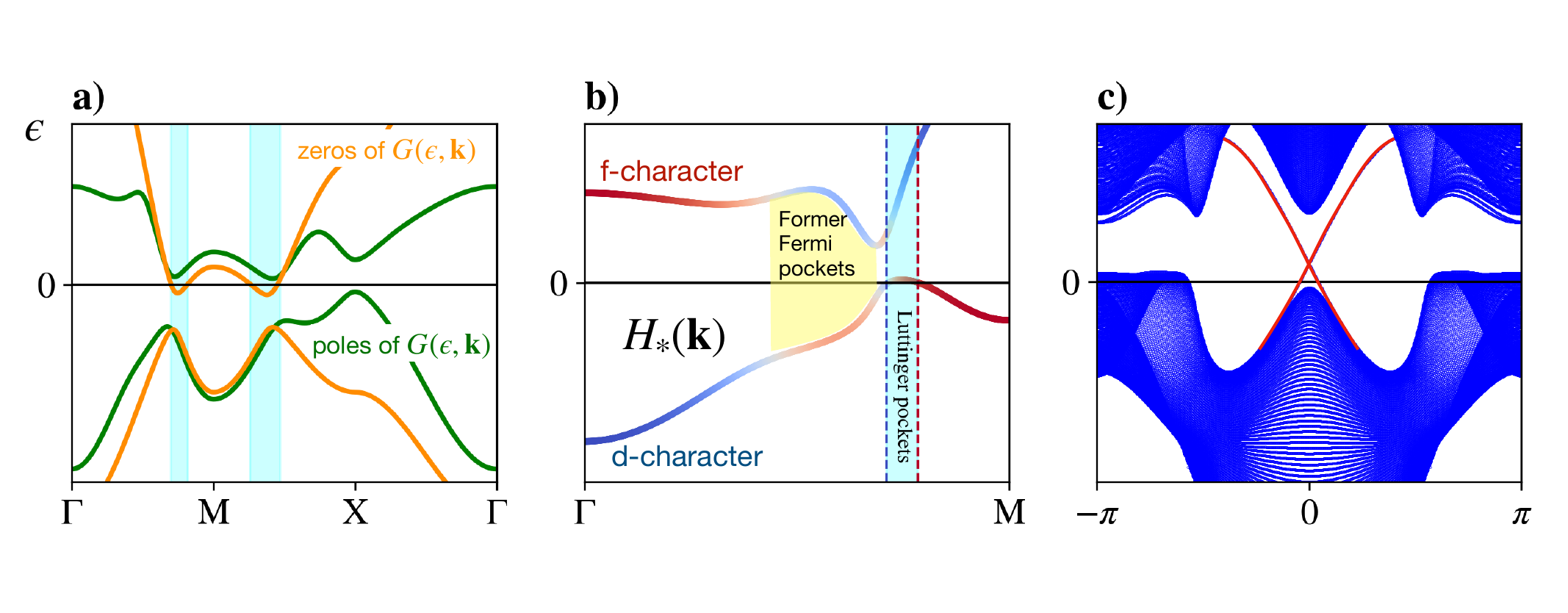}}
\caption{Panel (a): Poles and zeros of the eigenvalues of the Green's function $G(\ep,\bk)$ for real frequencies with the self-energy approximated as in \eqn{self-energy-ansatz} without particle-hole symmetry. The parameters are such that the bands of zeros cross the chemical potential, i.e., zero single-particle energy. The resulting Luttinger pockets are highlighted in cyan. Panel (b): Quasiparticle band structure of the model with PBC along $\bd{\Gamma}\to\mathbf{M}$. The Luttinger pocket is again highlighted in cyan. We also show in yellow the former Fermi pockets that are gapped by the inter-orbital hybridisation. Panel (c): Same as panel (b) but with OBC. We note, close to the zone boundaries, the bulk Luttinger pockets that cross the chemical potential. In addition, there are chiral edge states, in red, within the gap.
}
\label{figure3}
\end{figure*}

\noindent
The above results were obtained assuming that $f$ and $d$ orbitals are connected to each other by a particle-hole transformation, which greatly simplifies the numerical calculations. This symmetry inevitably causes the gapping of the Luttinger surfaces, but is not applicable to SmB$_6$ and YbB$_{12}$. A more realistic non-interacting Hamiltonian can be defined as the sum 
of the particle-hole symmetric one \eqn{H0_ph} plus a term~\cite{Assad1}
\beal
\delta H_0(\bk) &= \big(\eta\,\ep(\bk)-\mu\big)\,\tau_0\,,\label{delta H_0} 
\eal
with $0<\eta\lesssim 1$, which implies that the $f$ and $d$ hopping amplitudes are $t_f= (\eta-1)\,t$ and $t_d=(\eta+1)\,t$, respectively. If we suppose that also away from particle-hole symmetry 
the ansatz \eqn{self-energy-ansatz} with $\tilde{\Sigma}(\bk)=\Sigma^{HF}$ 
still approximates the true self-energy at low frequency, we can, upon tuning the 
parameters of $H_0(\bk)+\delta H_0(\bk)$ and of \eqn{self-energy-ansatz}, obtain the results shown in Fig.~\ref{figure3}. 
Specifically, in Fig.~\ref{figure3}(a) we plot the bands of poles and zeros of the eigenvalues of 
the Green's function $G(\ep,\bk)$ for real frequencies. We note that while the poles still describe 
an insulating state, a Luttinger surface survives because of the particle-hole asymmetry. This is 
also evident in the quasiparticle band structure of Fig.~\ref{figure3}(b) drawn along $\bd{\Gamma}\to\mathbf{M}$, which shows a \textit{quasiparticle Fermi surface} that is the Luttinger one. Remarkably, the quasiparticle spectrum in OBC, Fig.~\ref{figure3}(c), 
does exhibit edge modes, poles of the surface Green's function, which cross the chemical potential, 
despite the existence of the bulk \textit{quasiparticle Fermi surface}. This occurs since the bulk quasiparticles 
have vanishing residue and therefore cannot provide a decay channel for the edge modes.\\ 

\noindent
Figure~\ref{figure3} does realise the desired physical scenario that could rationalise 
the phenomenology of SmB$_6$ and YbB$_{12}$. Unfortunately, uncovering all such rich behaviour in a full DCA 
calculation is numerically unfeasible. However, we can still hope to find evidence in support. 
To this end, we consider the non-interacting Hamiltonian \cite{Assad1} $H_0(\bk)$ in \eqn{H0_ph} plus   
$\delta H_0(\bk)$ in \eqn{delta H_0}, setting $t=0.6$ and $\eta=0.4$, so that 
$t_d=1$ and $t_f=-0.2$, as well as $\lambda=0.4$ and $M=3$, while fixing $\mu$ so as to enforce 
$n_f+n_d=2$. 
The effect of the Coulomb repulsion is taken into account by its projection onto the local $d$ and $f$ orbitals of the rare earths, leading to the sum of three charge repulsion terms
\beal
H_U = \, \sum_\bR  \,\Big[ \,&U_f\, n_{f\bR\uparrow}\,n_{f\bR\downarrow}  + U_d\, n_{d\bR\uparrow}\,n_{d\bR\downarrow} + V \, n_{f\bR} \, n_{d\bR} \, \Big]\,,
\label{monopole}
\eal
along with a dipole-dipole exchange    
\be
H_J = -J\,\sum_\bR\,\bigg[\Big(f^\dagger_{\bR\up}\,f^\dagger_{\bR\down}\,d^\dagga_{\bR\down}\,
d^\dagga_{\bR\up} + H.c.\Big) + \sum_\sigma\,n_{f\bR\sigma}\,n_{d\bR\sigma}\bigg] 
 \, ,
\label{dipole}
\ee
which exists since the orbitals have opposite parity and enforces 
Hund's rules favouring high $|S_z|$ configurations. 
Estimating the relative strength between the Slater integrals using realistic effective charges for the atomic orbitals of Samarium~\cite{effective_charges}, we find $U_d=0.33\,U_f$, $V=0.46\,U_f$ and $J=0.1\,U_f$, ratios that we assume hereafter \footnote{Due to the 
complex nature of the auxiliary impurity model, here 
involving both $d$ and $f$ orbitals, 
we can only afford two momentum patches in DCA, one patch centred 
around $\bd{\Gamma}$ and the other around $\mathbf{M}$ \cite{Lorenzo-patches-PRB2009}. This choice,  
dictated by the increased numerical complexity, has the drawback of a limited momentum resolution. Unless stated otherwise, the numerical results are again obtained at $\beta=50$.}.\\

\begin{figure*}
\centerline{\includegraphics[width=1.0\textwidth]{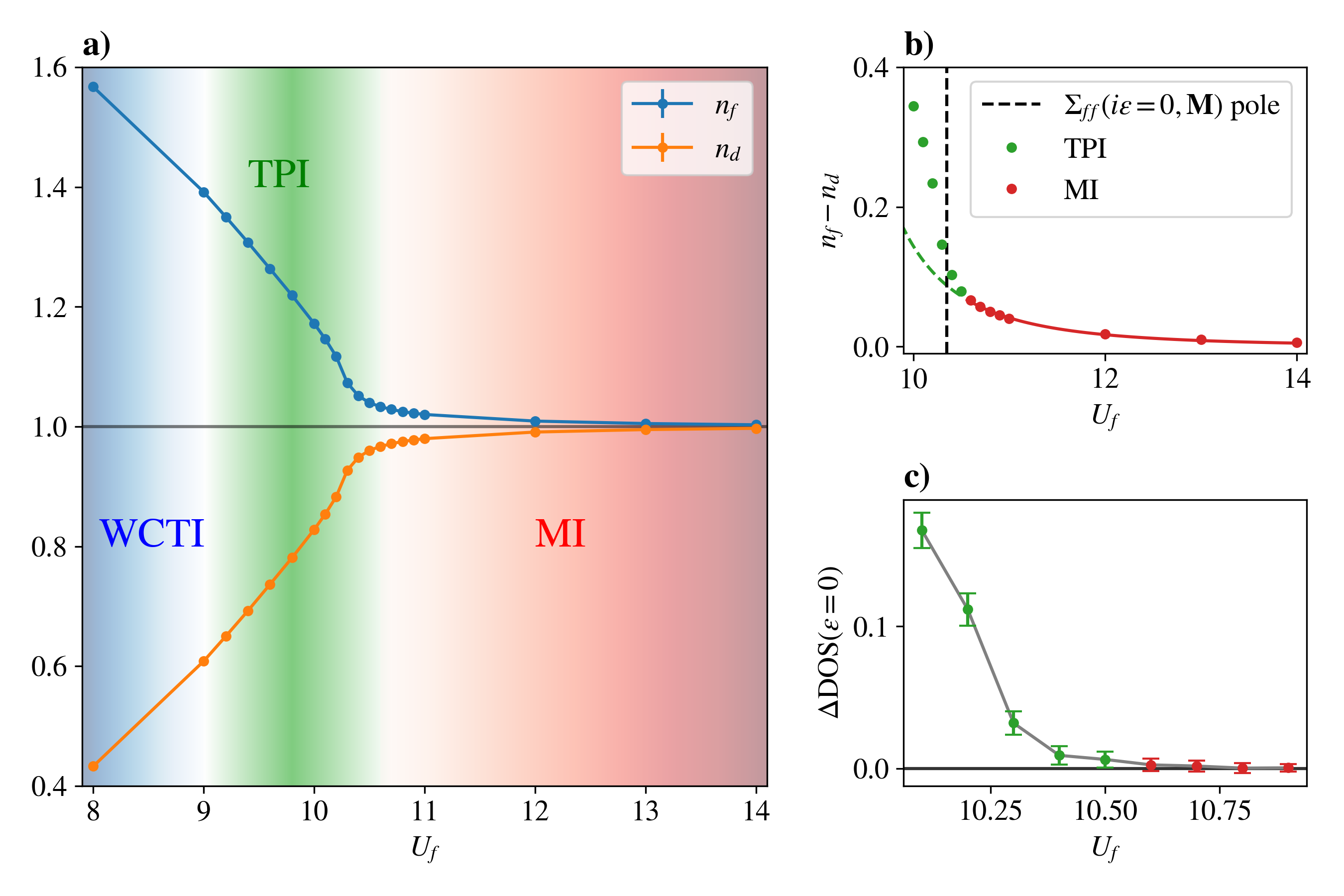}}
\caption{
Panel (a): Occupation numbers $n_f$ and $n_d$. At small $U_f$, the system describes a weakly correlated topological insulator (WCTI) adiabatically connected to the non-interacting one, while for large values of the interaction the system is a Mott insulator (MI). For intermediate values, a topological pseudogap insulator (TPI) emerges. 
Panel (b): Deviance of the orbital occupation from the $1/U_f^3$ dependence expected for large interactions in the MI regime, fitted on data for large $U_f$. We note that the numerical results deviate from the extrapolated behaviour when $U_f < 10.6$, hinting at a phase different from the MI. 
We also show the position of the $f$ self-energy pole at zero Matsubara frequency and at $\mathbf{M}$ 
that occurs when $U_f \sim 10.4$, see also Fig.~\ref{figure5}, testifying the presence of a Luttinger surface before the Mott transition.
Panel (c): Difference between the local density of states at the chemical potential without and with the inter-orbital hybridisation. This quantity is computed from the discontinuity at zero Matsubara frequency of $-\text{Im}G(i\ep,\bk) /2\pi$ integrated over $\bk$, and vanishes above $U_f \simeq 10.6$, consistently with the previous estimate.}
\label{figure4}
\end{figure*}

\noindent
In Fig.~\ref{figure4}(a) we show the orbital occupations $n_f$ and $n_d$ for increasing values of $U_f$. 
We note that $n_f$ is always greater than $n_d=2-n_f$ \cite{Giorgio-PRB2013}, which makes it hard to locate the Mott transition. However, deep in the Mott 
insulating phase one expects in perturbation theory that $n_f-n_d$ decays as $1/U_f^3$. 
We can exploit this observation to locate the onset of the Mott insulator at $U_{cf} \sim 10.6$, see 
Fig.~\ref{figure4}(b). An alternative estimate is obtained by noticing that in a genuine Mott insulator the spectral weight at the chemical potential is suppressed by strong correlations rather than by the inter-orbital hybridisation.  
Following this idea, we calculated the local density of states (DOS) at the chemical potential employing the converged cluster self-energies but setting $\lambda=0$ in the calculation of the local Green's function. In Fig.~\ref{figure4}(c) we present the difference between this quantity and the original DOS, non-zero because of the small but finite temperature. Also in this case the numerical results seem to indicate that the Mott transition occurs around $U_{cf}\sim 10.6$, when that difference reaches zero. 
\\

\begin{figure*}
\centerline{\includegraphics[width=1.0\textwidth]{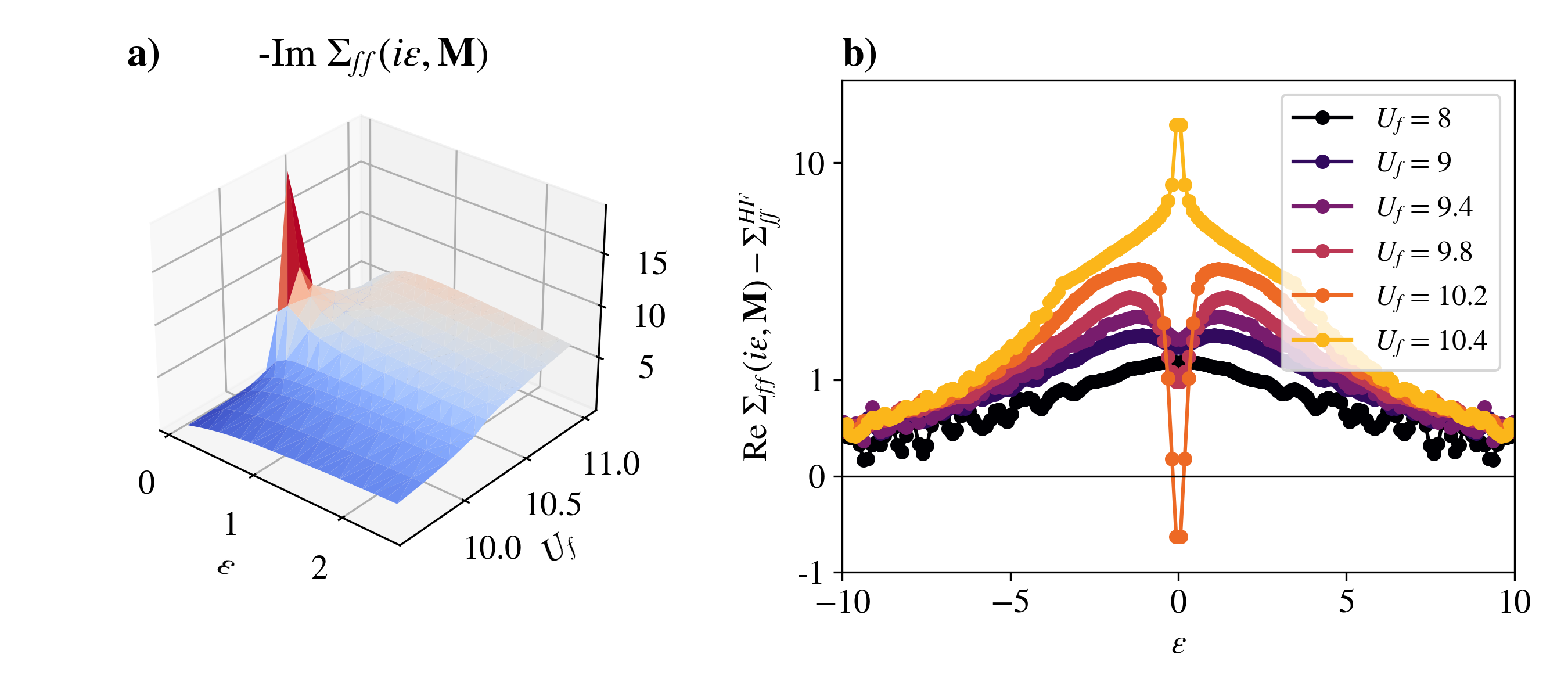}}
\caption{Panel (a): Behaviour of $-\Ima\Sigma_{ff}(i\ep,\bM)$ as function of $\ep>0$ and $U_f$. We note the appearance of a peak at $U_f=10.4$. 
Panel (b): $\Rea\Sigma_{f f}(i \ep, \mathbf{M} ) - \Sigma^{HF}_{f f}$ as function of $\ep$ and for different values of $U_f <U_{cf}$. We note a low frequency structure, now much more pronounced than in Fig.~\ref{figure2}(a), which we 
associate with a Luttinger surface that exists in the patch and which 
seems to cross the point $\bM$ when $\Ima\Sigma_{ff}(i\ep,\bM)$ develops a singularity.}
\label{figure5}
\end{figure*}

\noindent
Strikingly, the system shows compelling indications of strong correlations even before $U_{cf}$. In Fig.~\ref{figure5}(a) we plot $\Ima\Sigma_{ff}(i\ep,\bM)$ as function of the Matsubara frequency $\ep$ and for different values of $U_f$. We observe a singular behaviour around $U_f=10.4$, thus before $U_{cf}$ and within the topological insulating phase, signalling the proximity of a Luttinger surface. We note that because of the limited $\bk$-space resolution, $U_f=10.4$ is just an upper estimate, since the Luttinger surface may well emerge earlier but away from $\bM$. 
Similarly to Fig.~\ref{figure2}(a), we plot in Fig.~\ref{figure5}(b) $\Delta\Sigma_{f f}(i \ep, \mathbf{M} )= \Rea\Sigma_{f f}(i \ep, \mathbf{M} ) - \Sigma^{HF}_{f f}$, as \textcolor{blue}{a} function of $\ep$ and for different values of $U_f <U_{cf}$.  
In the weakly correlated topological insulator, $U_f \lesssim 8$, this quantity is smooth and monotonically decreases in $|\ep|$. This behaviour stops at $U_f \sim 9$ where $\Delta\Sigma_{f f}(i \ep, \mathbf{M} )$ develops a minimum at $\ep=0$ that grows and sharpens until, around $U_f=10.4$ when $\Ima\Sigma_{ff}(i\ep,\bM)$ develops a singularity, a sharp maximum. That sudden change of behaviour at $U_f\gtrsim 9$, narrowly centred around $\ep=0$, we believe signals the appearance of a Luttinger surface within the patch, which crosses the $\bM$ point at $U_f$ around $10.4$.  \\

\begin{figure*}
\centerline{\includegraphics[width=1.0\textwidth]{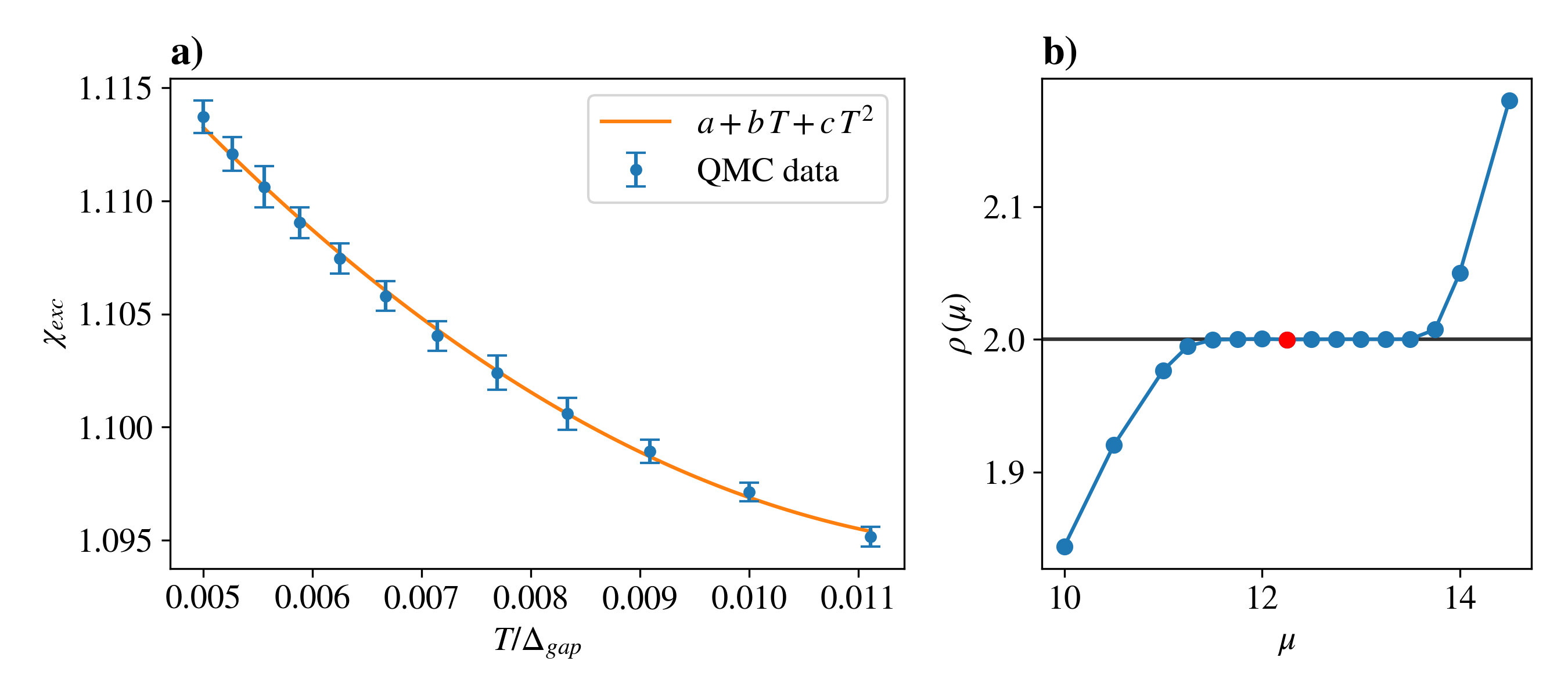}}
\caption{Panel (a): Susceptibility to a field that couples to the odd-parity particle-hole excitations as function of temperature, in units of the gap, deep in the Mott phase, $U_f=16$ and at $\mu=12.25$, see Supplementary Material. The data, obtained for $ 45 \leq \beta \leq 100$, show an excellent agreement with a polynomial fit up to a quadratic term in temperature.
Panel (b): Behaviour of the system density upon varying the chemical potential at $U_f=16$ and $\beta=100$, showing a clear plateau compatible with a sizeable insulating gap $\Delta_{gap} \sim 2 $. The red dot corresponds to the chemical potential employed in the calculations of panel (a).}
\label{figure6}
\end{figure*}

\noindent
One remaining question is the actual existence of gapless neutral quasiparticles associated with the Luttinger surface. The easiest way to infer their presence is by studying the response to external fields. The most natural choice would be the magnetic response \cite{Sawatzky-PRR2023}. 
Unfortunately, that is known to be severely suppressed in two-patch and four-patch DCA, owing to the formation of local singlets inside the impurity cluster \cite{Maier-PRL2005}, as we indeed find. Therefore, we 
focus here on a different excitation channel that 
is also important in connection with the anomalous optical conductivity of SmB$_6$ \cite{Armitage-PRB2016}. Specifically, we consider the response to a field that couples to the odd-parity particle-hole 
excitation $\propto \tau_2\,\sigma_1$, a channel which has the most attractive \textit{bare} scattering amplitude  
\cite{Blason2020PRB}.
We note that at $T=0$ the finite response to this field cannot be used as hallmark of the existence 
of in-gap excitations because the operator is not a conserved quantity. The temperature 
dependence, on the other hand, is a discriminating factor since in fully-gapped insulators we expect an activated behaviour, 
$\sim \exp(-\Delta_{gap}/T)$, with $\Delta_{gap}$ the insulating gap, while a power-law behaviour 
in presence of gapless excitations.
\\
Fig.~\ref{figure6}(a) shows the  
static uniform susceptibility as function of $T$ at $U_f = 16$ deep in the Mott regime, where we are able to directly observe the divergence of the cluster self-energy inside the gap, see Supplementary Material. In this case we do find a power-law behaviour, which is striking since the model has a sizeable charge gap, see Fig.~\ref{figure6}(b). 
We emphasise that for the reported values of $\Delta_{gap}/T\in[90,200]$, an activated 
behaviour $A - B\exp(-\Delta_{gap}/T)\simeq A$ cannot fit in any way the data  
\footnote{In the Supplementary Material we also report the response in the TPI phase, where, however, Monte Carlo errors are too large and the ratio between temperature and gap too high to discern a power-law dependence on temperature.
In any case, the presence of gapless neutral excitations only requires a Luttinger surface \cite{Michele2,Michele3}, so we believe that the evidence obtained at $U_f=16$ can be extended to lower interaction values, and, in particular, to the TPI phase, as long as a Luttinger surface exists.}.\\

\noindent
In conclusion, we propose a physical scenario that explains the unconventional properties of 
SmB$_6$ and YbB$_{12}$ Kondo insulators.  
It assumes, in essence, 
topological valence and conduction bands separated by a hybridisation gap  coexisting with 
Luttinger surfaces. The bands account for the non-trivial topology that yields the edge modes 
responsible for the surface metallic character. The Luttinger surfaces, instead, may host quasiparticles that do not contribute to electric transport, but determine  thermal transport, quantum oscillations in  magnetic field \cite{Michele3} and, possibly,  in-gap optical absorption. Unveiling this rich phenomenology has proved possible even in DCA calculations, thanks to intense numerical work. Several points remain open for further study. Our initial assumption was that SmB$_6$ and YbB$_{12}$ are close to a Mott insulating phase,  
identified with the magnetic 
$4f^{n-1}\,5d^1$ configuration, $n=6$ and $n=14$ for Sm and Yb, respectively. This assumption 
is likely valid in YbB$_{12}$ \cite{Mixed-valence-Yb} but more questionable in SmB$_6$ \cite{Mixed-valence-Sm}. 
On the other hand, it is still possible that the $4f^6 d^0$ configuration would also corresponds to 
a Mott insulator, this time non-magnetic. That would imply that SmB$_6$ is halfway between two 
Mott phases and, presumably, that our topological pseudogap regime is even broader 
than in YbB$_{12}$. Even though this physics goes beyond our simple modelling with single $f$ and $d$ orbitals, we believe that the overall qualitative scenario would remain the same. \\
Another crucial aspect is that the pseudogap phase in all model calculations, including ours, is restricted to a narrow region of 
Hamiltonian parameter space. Therefore, it may seem surprising that both SmB$_6$ and YbB$_{12}$ happen to lie precisely in that region. That being said, our model  evidently provides an extremely simplified cartoon of those materials. Yet we believe one which effectively illustrates the interplay between poles and zeros of the Green's function in the pseudogap phase. It also emphasises how Fermi liquid theory at a Luttinger surface \cite{Michele2,Michele3,libro} can be regarded as a semi-phenomenological theory just like that at a Fermi surface \cite{Landau1}. In that way, the Luttinger surface scenario that we propose is legitimate in so far it provides a consistent framework for 
the phenomenology of SmB$_6$ and YbB$_{12}$, as we believe it does 
\footnote{We remark \cite{Michele3} that a Luttinger surface just like a Fermi surface is likely to become unstable at sufficiently low temperature. What is important for the Fermi liquid properties to be observable is only that the quasiparticle degeneracy temperature exceeds the instability one.}.

\bmhead{Methods}
The results we presented were obtained using the so-called dynamical cluster approximation (DCA) \cite{RevModPhys.77.1027}, which consists in partitioning the Brillouin zone into patches and in approximating the self-energy as a piecewise constant function within these patches. The many-body problem is tackled by self-consistently solving coupled quantum impurity models. The impurity solver is implemented within the TRIQS library \cite{PARCOLLET2015398} and makes use of a continuous time quantum Monte Carlo algorithm based on a hybridization expansion of the partition function \cite{SETH2016274}. \\
For the 4-patches computations we used four squared patches centred around the high-symmetry points $\mathbf{\Gamma}$, $\mathbf{X}$, $\mathbf{Y}$ and $\mathbf{M}$. In the 2-patches computations instead, we employed just two patches centred around $\mathbf{\Gamma} $ and $\mathbf{M} $, since four patches were numerically not affordable. Specifically, we used the same partition of the Brillouin zone as in \cite{Lorenzo-patches-PRB2009}, in which the patch at $\bM$ includes also the $\mathbf{X}$ and $\mathbf{Y}$ points. \\
The results shown in OBC were obtained starting from the periodic dispersion of the poles/zeros as a function of $k_x$, $k_y$ and constructing the matrices of a slab with OBC along y with $N_y = 200$ layers and PBC in the $x$ direction.

\bmhead{Acknowledgements}
We are very grateful to Antoine Georges, Elio K\"{o}nig, Giorgio Sangiovanni and, especially, to Adriano Amaricci, Carlos Mejuto Zaera and Erio Tosatti for helpful discussions and comments.


\end{document}


\title{Supplementary material to "Luttinger surface dominance and Fermi liquid behaviour of topological Kondo insulators  SmB$_6$ and YbB$_{12}$"}

\author[1]{Andrea Blason}\email{ablason@sissa.it}
\author[1]{Ivan Pasqua}\email{ipasqua@sissa.it}
\author[2,3]{Michel Ferrero}\email{michel.ferrero@polytechnique.edu}
\author[1]{Michele Fabrizio}\email{fabrizio@sissa.it}

\affil[1]{International School for Advanced Studies (SISSA), Via Bonomea 265, I-34136 Trieste, Italy} 

\affil[2]{CPHT, CNRS, Ecole Polytechnique, Institut Polytechnique de Paris, Route de Saclay, 91128 Palaiseau, France}
\affil[3]{Coll\'ege de France, 11 place Marcelin Berthelot, 75005 Paris, France} 

\maketitle
\clearpage
\section{Additional results for the particle-hole symmetric model}

\begin{figure*}[h!]
\centering
\includegraphics[width = 1.0 \textwidth]{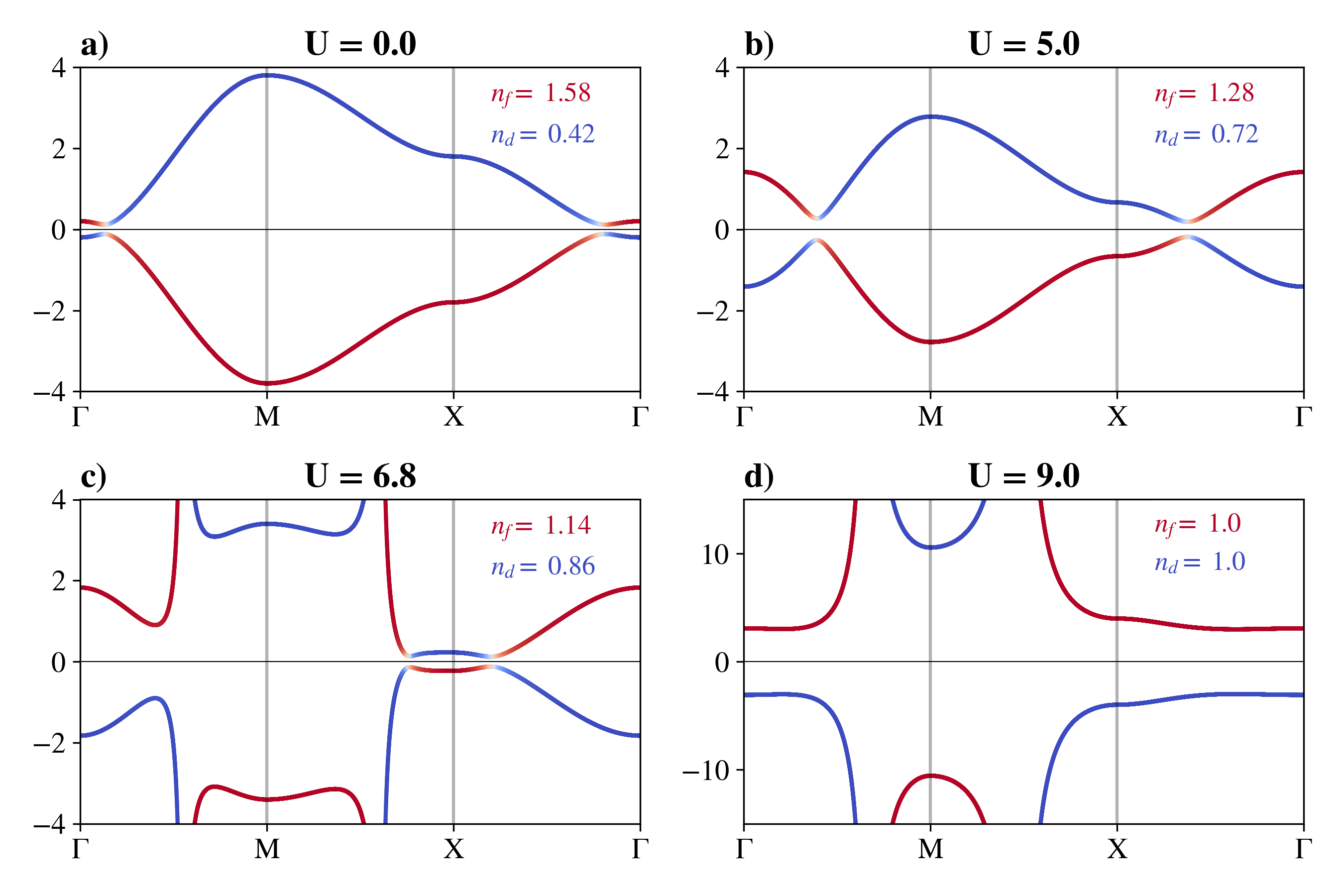}
\caption{\textbf{ Renormalized energies $\mathbf{ R(\mathbf{k}) = H_{0}(\mathbf{k} )+ \text{Re} \Sigma(0, \mathbf{k}) }$ across the Quantum Spin Hall Insulator (QSHI) to Mott Insulator (MI) transition for $\mathbf{ M = 1.8 }$ and four values of $\mathbf{U}$.}  The color plot represents the orbital flavour of the eigenvectors. $n_{f}$ and $n_{d}$ are the $f$ and $d$ occupation numbers, respectively. Panel a): At $U = 0$ the model is a non-interacting QSHI with band inversion close to $\bd{\Gamma}$ and a topological hybridisation gap. Panel b): At $U=5$ the renormalised bands are still adiabatically 
connected to the non-interacting ones. Panel c): AT $U=6.8$, the two orbitals are close to half-filling and one recognises all signatures of the Topological Pseudogap Insulator (TPI): the Luttinger surfaces, 
poles of the $R(\bk)$ eigenvalues, and gapped Fermi pockets around $\mathbf{X}$. The Chern number of the valence band is the same of the non-interacting one, $C_{\uparrow} = - C_{\downarrow} = 1 $. Panel d): 
At large $U=9$, $R(\bk)$ describes a MI: both orbitals are half-filled, at least within our numerical accuracy, the topological invariant \cite{Andrea-PRB2023} is only carried by the bands of Green's function zeros and its value is the opposite of that in the QSHI and TPI insulators. }
\label{fig-Rk_M3}
\end{figure*}

\clearpage

\section{Additional results for the particle-hole asymmetric model}
\begin{figure*}[h!]
\centering
\includegraphics[width = 0.9 \textwidth]{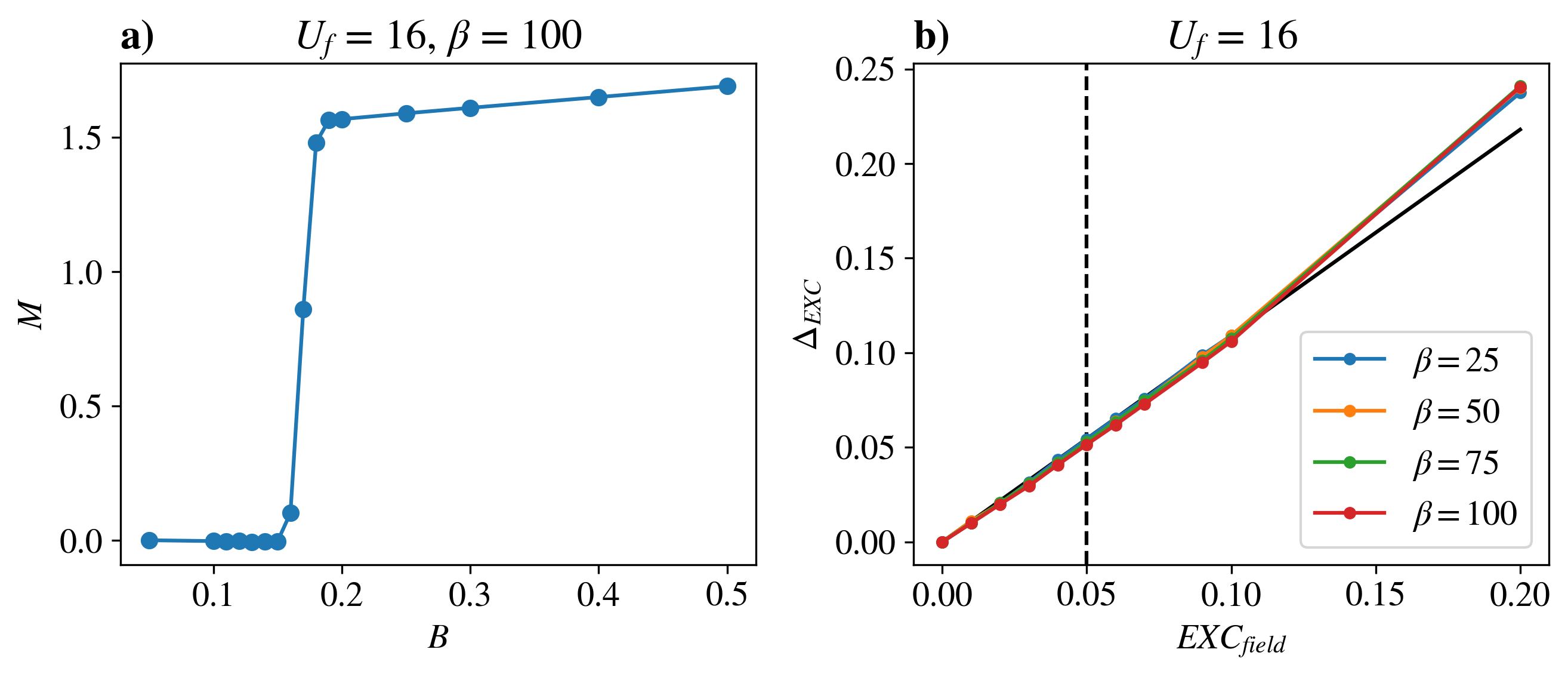}
\includegraphics[width = 0.9 \textwidth]{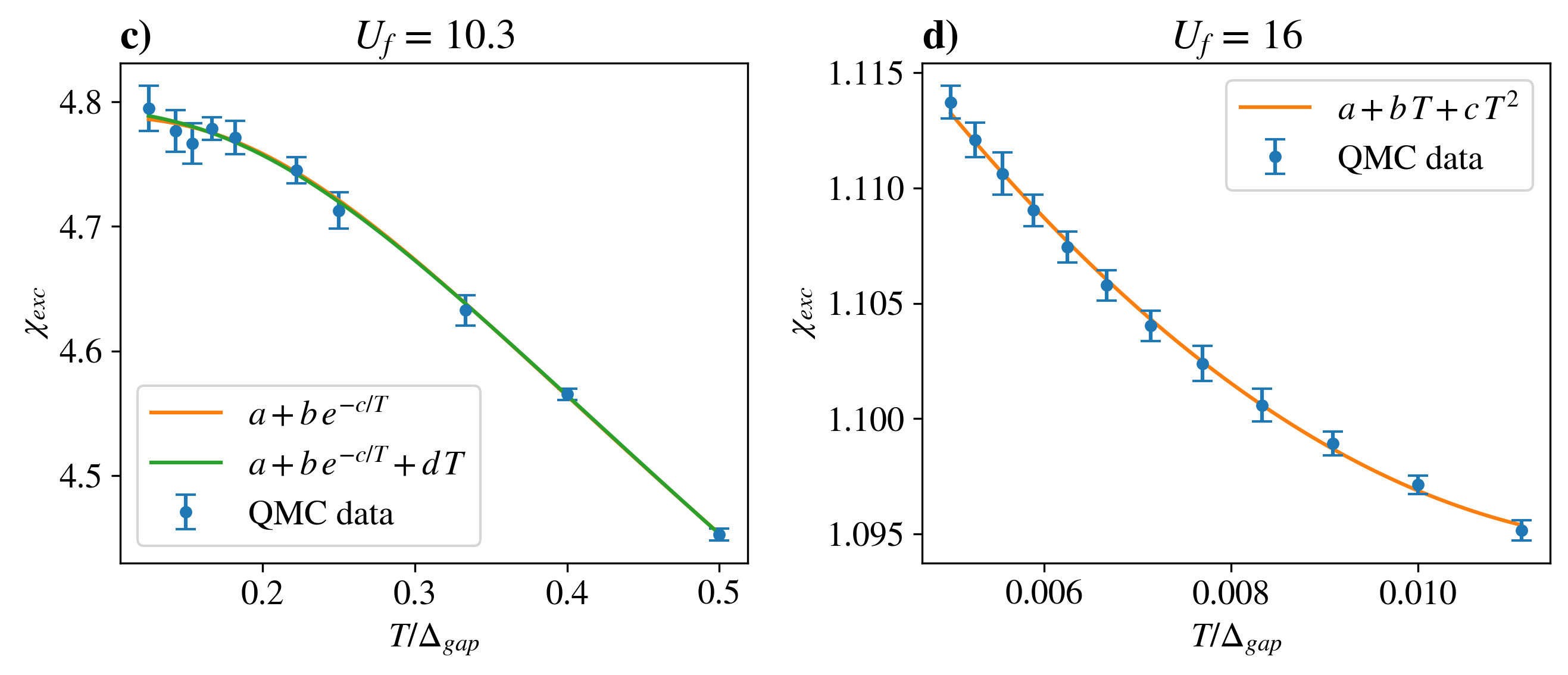}
\caption{\textbf{Panel a) and b): expectation values of operators vs. the external fields 
to which they are coupled deep in the Mott phase.} Panel a) : Response to a static and uniform magnetic field: up to a critical value $B_c \sim 0.15$, the system is essentially non-magnetic because the impurity cluster locks into a spin-singlet, which is an artefact of the two-patch DCA. Panel b): The same as panel a) but for the excitonic odd-parity operator $\Delta_{EXC}\sim \tau_2 \sigma_1$ that is most favoured by interaction and for  different temperatures. The vertical dashed line is the value of the field employed to calculate the linear part of the static response function, which in this case can be well approximated by $\chi_{ext} = \Delta_{EXC}/EXC_{field}$. The solid black line is just a guide to the eye to help identify the linear regime. 
\textbf{Panel c) and d): comparison of the linear term of the linear susceptibility to the field of panel b) for different values of $U_f$.} Panel c) : The response in the TPI phase is reported. Due to the sizeable Monte Carlo errors and the high temperatures relative to the gap $\Delta_{gap}$ we could access by the hybridisation-expansion impurity solver, it was not possible to discern the presence of a genuine linear in temperature dependence. In fact, see panel d), that would require going to much lower $T/\Delta_{gap}$ where the contribution of the neutral quasiparticle is expected to dominate.
}
\end{figure*}

\begin{figure*}
\centering
\includegraphics[width = 1.0 \textwidth]{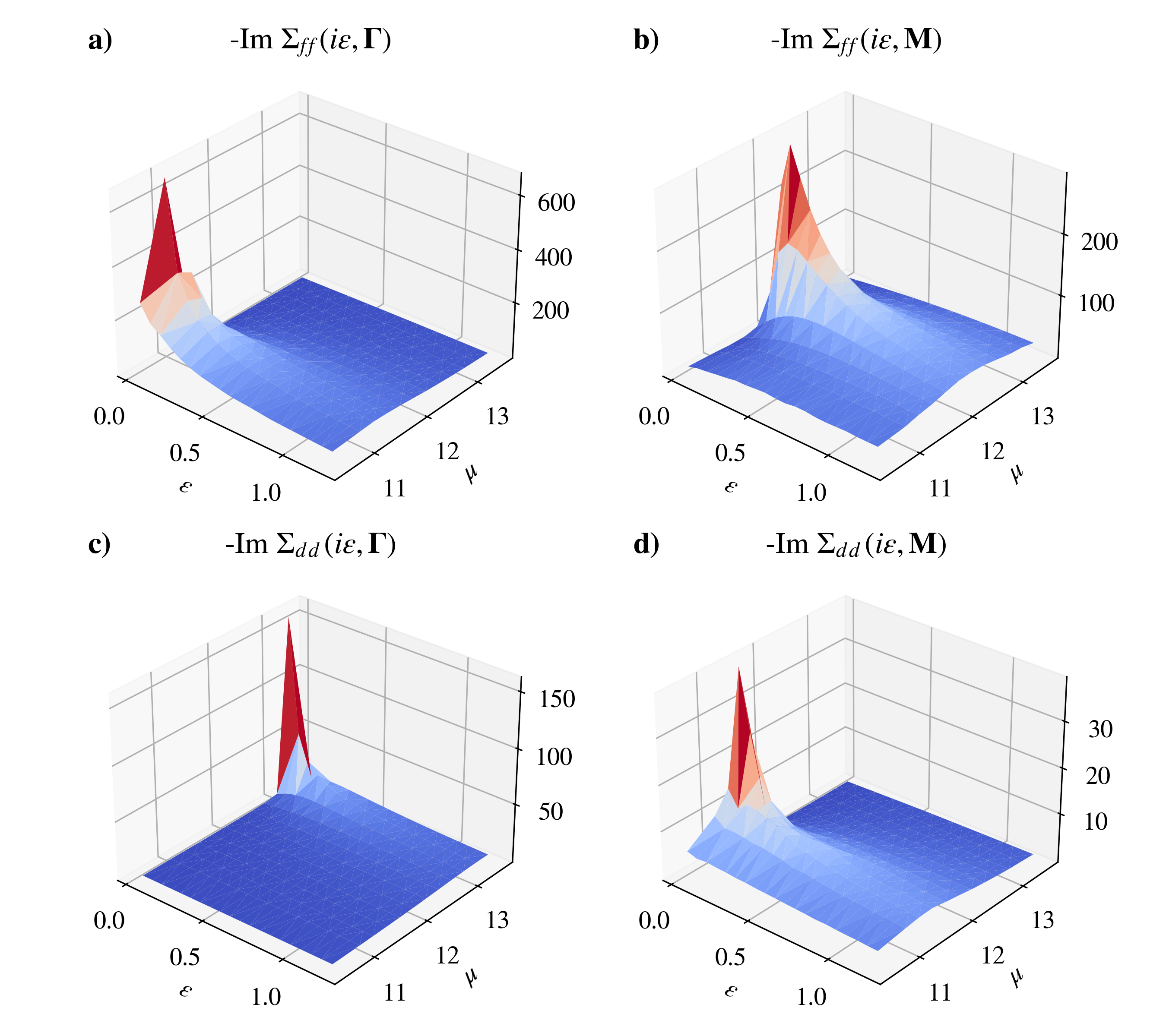}
\caption{Imaginary part of the cluster self-energies as function of the Matsubara frequency and chemical potential $\mu$ inside the gap, at $U_f=16$ and $\beta=100$. In this case, due to the sizeable insulating gap, true divergences of $\Ima\Sigma(i\ep,\bk)$ for $\ep\to 0$ can be directly observed without any periodisation scheme, simply by moving the chemical potential inside the gap. These divergences 
indicate a two-orbital Mott insulator with in-gap dispersive bands of Green's function zeros.}
\end{figure*}

\clearpage

\section{Topological periodic Anderson model}

\begin{figure*}
\centering
\includegraphics[width = 0.95 \textwidth]{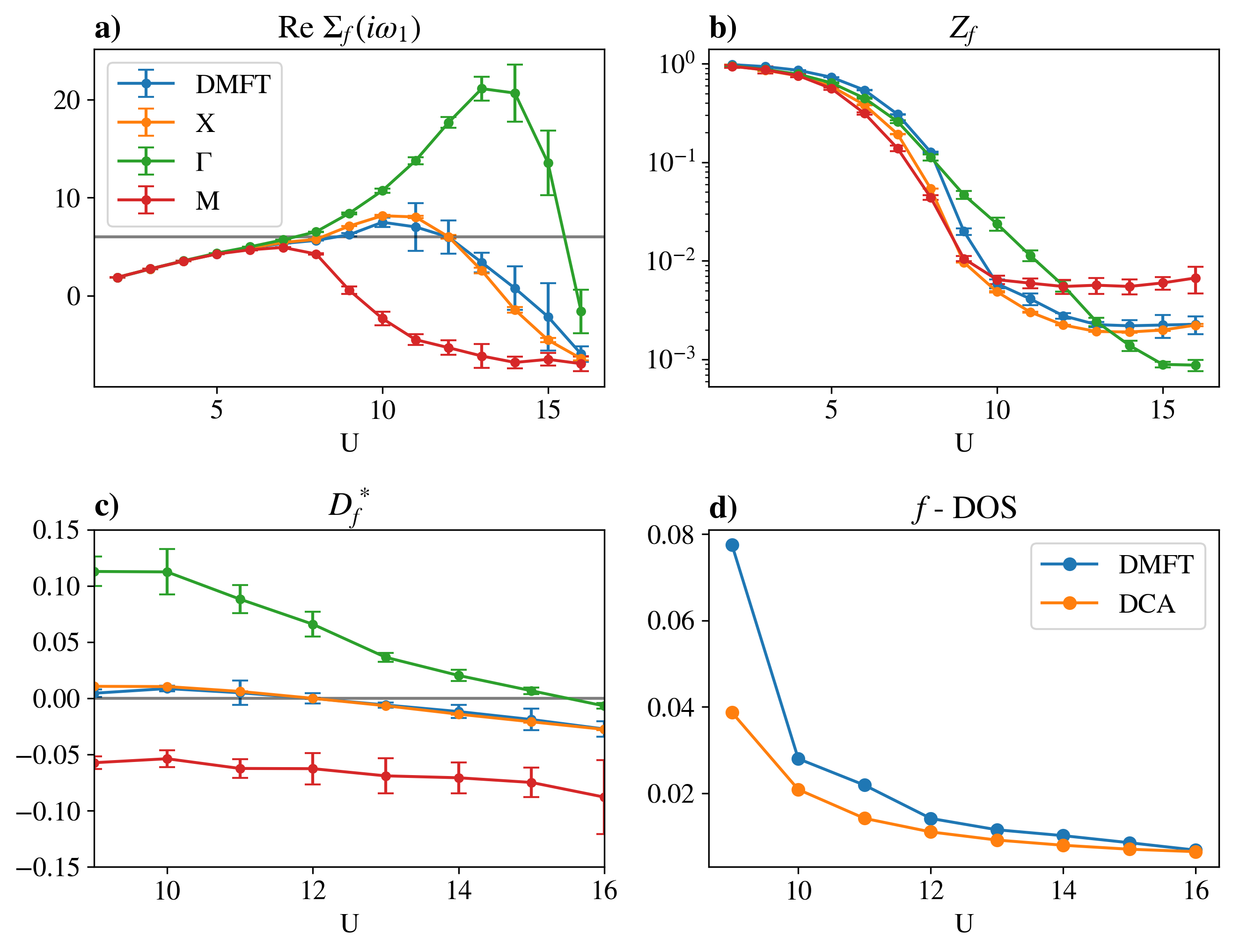}
\caption{\textbf{Numerical results for the topological periodic Anderson model}. The DMFT calculations has been performed at $\beta=100$, while the 4-patch DCA ones at $\beta=50$. In particular, the results of panels a), b) and c) are averaged over five self-consistent iterations and spin, with the error bars corresponding to the standard deviation. In panel d), on the other hand, the average over iteration has been performed prior to the fitting procedure. Moreover, since the $\bX$ and $\bY$ points are equivalent, only results for the former are presented. Finally, the color labelling is consistent in every panel.
Panel a): Real part of the $f$-self energy at the first Matsubara frequency $\omega_1=\pi T$. The grey line corresponds to $D_f + \text{Re}\Sigma_f = 0$, where the gap closes due to the vanishing hybridisation. 
Panel b): Quasiparticle residue of the $f$-band $Z_f = 1/m_f$ in logarithmic scale.
Panel c): Quasiparticle crystal field splitting 
$D_{*f}= Z_f\,\left(D_f + \text{Re}\Sigma_f(i\pi T)\right)$.
Panel d): $f$-density of states (DOS) at the chemical potential, estimated from the interpolation at zero frequency of the local Green's function. In particular, we fitted the Green's function imaginary part on the Matsubara axis around $\ep =0$ with a trial-function $ - a \, \text{sign}(\ep) + b \, \ep + c \, \ep^3 + d \, \ep^5$, so that the DOS corresponds to the estimated value $a / 2\pi $.}
\label{fig-TPAM1}
\end{figure*}
We complement the results presented in the main article by analysing the particle-hole asymmetric model, but now with only $U_f\neq 0$. The model, where now only the $f$-electrons are interacting, has been investigated \textit{via} dynamical mean field theory (DMFT) by Werner and Assaad~\cite{Assad1,Assad2}, whose findings are worth discussing. In this section we will try to uniform our notation to the one used by those authors to make the comparison easier. In particular we define the crystal field splitting as the difference between the bare on-site energy of the f and d orbital $D_f = - 2 M = -6.0$, in our case. Werner and Assaad showed that the main effect of interaction is a renormalization of the effective crystal field splitting (CFS) by the zero-frequency part of the $f$-self energy:
\be \label{eq-ECFS}
\widetilde{D}_f = D_f + \Rea\, \Sigma_f(i\pi T)  \, .
\ee
where the (real) zero-frequency value of the self-energy is approximated with the one at the first Matsubara frequency, directly accessible from imaginary time calculations. \\
Specifically, owing to its Hubbard nature, increasing the interaction shifts the $f$-band center to achieve half-filling and suppresses double occupations. Simultaneously, the effective $f$-mass (or the inverse of the quasiparticle residue):
\be \nonumber
m_f = Z_f^{-1} \equiv \left( 1 - \fract{\Ima \Sigma_f(i\pi T)}{\pi T} \right) \, ,
\ee
grows rapidly, pushing the $f$-electrons into the heavy-fermion regime and strongly suppressing the effective hybridisation
\be 
\lambda_* = \sqrt{Z_f\,} \; \lambda \, .
\ee
At $U \sim 9$, the effective mass appears to diverge, indicating that the $f$-electrons localise and turn 
into local moments almost decoupled from the $d$-electrons. The authors of \cite{Assad1,Assad2} argue  that, for larger interaction strengths, the model may undergo an orbital selective Mott transition (OSMT), though they do not delve further into the analysis of this scenario.

We complement these results by exploring this strongly interacting regime by means of both DMFT and by 
4-patch DCA, which allows the computation of the self energy at the four momenta $\bGamma=(0,0)$, $\bM=(\pi,\pi)$, $\bX=(\pi,0)$ and $\bY=(0,\pi)$, with the latter two equivalent by $C_4$. We stress that in DCA the effective crystal field splitting calculated at the $\bX$ and $\bY$ points, easily generalised from Eq.~\eqref{eq-ECFS}, is the actual parameter that controls the topological transition since the gap closes at these points when the $f$-band becomes half-filled.
\\
In the regime discussed in \cite{Assad1,Assad2} the physics is dominated by local correlations, so that DMFT and DCA self-energies calculated at the $\bX$ point essentially coincide for $U < 9$, see Fig.~\ref{fig-TPAM1}a. The inclusion of non-local corrections induces just a renormalisation of the 
short-range hopping, manifesting as a small deviance of the self-energy values at the different cluster momenta. 
\\
At $U \sim 9$ the effective CFS becomes small and the quasiparticle mass very large, see panels a) and b) of Fig.~\ref{fig-TPAM1}, both effects diminishing the gap that gradually moves towards the $\bX$ and $\bY$ points. 
However, upon further increasing $U$, we do not find any evidence of OSMT, 
as expected because of the finite inter-orbital hybridisation \cite{OSMT_Kotliar}.
Instead, we find that the effective CFS remains small in absolute value. 
At $U=12$, it even crosses zero, both in DMFT as well as at the $\bX$ and $\bY$ points in DCA. Here, the model effectively recovers particle-hole symmetry, the two bands cross at $\bX$ and $\bY$ where the hybridisation vanishes, and therefore the system is semimetallic. 
However, apart from this single value of $U$, we find a heavy-fermion behaviour with a hybridisation gap slightly offset from 
the $\bX$ and $\bY$ points and so tiny that thermal fluctuations can drive finite single-particle weight at the chemical potential, see Fig.~\ref{fig-TPAM1}d. 

These numerical results establish a clear physical scenario: for $U \lesssim 9$, the system behaves as a weakly correlated topological insulator with a sizeable gap, exhibiting a non-trivial $Z_2$ invariant and chiral edge states~\cite{Assad2}. As the interaction strength further increases, the $f$-band becomes flatter and flatter, it pins at the chemical potential and progressively loses spectral weight in favour of the Hubbard bands. Nonetheless, the model remains a 
topological Kondo insulator with a tiny gap that narrows down increasing $U$ and exactly vanishes at the single point $U = 12$.

The above results explicitly show that $U_f$ alone cannot explain the phenomenology of SmB$_6$ 
and YbB$_{12}$. 

\clearpage

\section{Few remarks about Landau's Fermi-liquid theory}
A natural concern about our Luttinger surface scenario is how it is possible that the quasiparticles 
do not contribute to the charge transport, as if they were neutral, and yet they are coupled to a 
magnetic field and yield quantum oscillations. Here, we aim to clarify this issue within the 
conventional Landau Fermi-liquid theory \cite{Pines&Nozieres,libro}. \\
We recall the Landau-Boltzmann transport equation in absence of collisions and for the charge 
component $\delta n_\bk(t,\br) = \delta n_{\bk\up}(t,\br)+\delta n_{\bk\down}(t,\br)$ of the deviation from equilibrium of the quasiparticle distribution 
function, 
\beal
\fract{\partial \delta n_{\bk}(t,\br)}{\partial t} + \bd{v}(\bk)\cdot\bd{\nabla}_\br\,
\overline{\delta n}_{\bk}(t,\br) -\fract{e}{c}\;\bd{v}(\bk)\wedge\bd{B}\cdot
\bd{\nabla}_\bk\,\overline{\delta n}_{\bk}(t,\br) = \fract{2e}{V}\;
\fract{\partial f\big(\ep_*(\bk)\big)}{\partial\ep_*(\bk)}\;\bd{v}(\bk)\cdot\bd{E}(t,\br)\,,
\label{LBEq}
\eal
where $\ep_*(\bk)$ is the quasiparticle dispersion measured with respect to the chemical potential \footnote{Eq.~\eqn{LBEq} implicitly assumes that the quasiparticles do not possess a Berry curvature \cite{PhysRevLett.93.206602,PhysRevLett.109.181602}.}, 
\beal
\bd{v}(\bk) = \fract{\partial\ep_*(\bk)}{\partial\bk}\;,
\label{def: group velocity}
\eal
the quasiparticle group velocity, $\bd{B}$ a static and uniform magnetic field, 
$\bd{E}(t,\br)$ the electric field, and, finally, 
\bealn
\overline{\delta n}_{\bk}(t,\br) \equiv 
\delta n_{\bk}(t,\br) -\fract{\partial f\big(\ep_*(\bk)\big)}{\partial\ep_*(\bk)}\;
\fract{2}{V}\,\sum_{\bk'}\,f_{S\,\bk\bk'}\,\delta n_{\bk'}(t,\br)\,,
\eal
is the deviation from local equilibrium, i.e., when the quasiparticle at $\bk$ 
is in equilibrium with other excited quasiparticles \cite{Pines&Nozieres}. \\
When $\bd{B}=\bd{E}=0$, Eq.~\eqn{LBEq} provides the definition of the charge current via the continuity equation
\beal
\bd{J}(\br) &= -e\,\fract{1}{V}\,\sum_\bk\,\bd{v}(\bk)\,
\overline{\delta n}_{\bk}(\br)
= -e\,\fract{1}{V}\,\sum_\bk\,\overline{\bd{v}}(\bk)\,
\delta n_{\bk}(\br)\,,\label{current}
\eal
where 
\beal
\overline{\bd{v}}(\bk) &= \bd{v}(\bk)
-\fract{\partial f\big(\ep_*(\bk)\big)}{\partial\ep_*(\bk)}\;
\fract{2}{V}\,\sum_{\bk'}\,f_{S\,\bk\bk'}\,\bd{v}(\bkp)\,,
\label{transport current}
\eal
is the charge-transport velocity. 
In presence of a uniform electric field $\bd{E}(t,\br)=\bd{E}(t)$ at $\bd{B}=0$, 
Eq.~\eqn{LBEq} becomes simply 
\bealn
\fract{\partial \delta n_{\bk}(t)}{\partial t} = \fract{2e}{V}\;
\fract{\partial f\big(\ep_*(\bk)\big)}{\partial\ep_*(\bk)}\;\bd{v}(\bk)\cdot\bd{E}(t)\,,
\eal
thus
\bealn
\fract{\partial\bd{J}(t)}{\partial t} &= -\fract{2e^2}{V}\;\sum_\bk\,
\fract{\partial f\big(\ep_*(\bk)\big)}{\partial\ep_*(\bk)}\;
\overline{\bd{v}}(\bk)\,\Big(
\bd{v}(\bk)\cdot\bd{E}(t)\Big)
=-\fract{2e^2}{V}\;\sum_\bk\,\overline{\bd{v}}(\bk)\,
\Big(\bd{E}(t)\cdot\bd{\nabla}_\bk\,f\big(\ep_*(\bk)\big)\Big)\\
&= \sum_{i=1}^d\,\fract{2e^2}{V}\;\sum_\bk\,f\big(\ep_*(\bk)\big)\;
\partial_{k_i}\overline{\bd{v}}(\bk)\;E_i(t)\,,
\eal
where $d>1$ is the system dimension. Therefore, the retarded component of the 
Fourier transform in time reads 
\beal
J_i(\omega) &= \fract{i}{\;\omega+i0^+\;}\,\sum_{j=1}^d\,\fract{2e^2}{V}\;\sum_\bk\,f\big(\ep_*(\bk)\big)\;
\partial_{k_j}\overline{v}_i(\bk)\;E_j(\omega)
\equiv \sum_{j=1}^d\,\sigma_{ij}(\omega)\;E_j(\omega)\,,
\label{current vs E}
\eal
which defines the conductivity tensor in absence of collisions. Its diagonal components, which we assume 
isotropic, directly yield the Drude weight
\beal
D_* &= \fract{\;\omega+i0^+\;}{i}\,\sigma_{ii}(\omega) = \fract{2e^2}{V}\;\sum_\bk\,f\big(\ep_*(\bk)\big)\;
\partial_{k_i}\overline{v}_i(\bk) 
= \fract{2e^2}{dV}\;\sum_\bk\,f\big(\ep_*(\bk)\big)\;
\bd{\nabla}_\bk\cdot\overline{\bd{v}}(\bk) \,.\label{Drude}
\eal
Our aim is to show that \eqn{LBEq} is consistent with the system being a Mott insulator. This necessarily implies that the Drude weight \eqn{Drude} must vanish, which is the case if the flux of $\overline{\bd{v}}(\bk)$ out of the 
quasiparticle Fermi surface is zero. We emphasise that this requirement is perfectly consistent with a finite group velocity $\bd{v}(\bk)$, and thus 
with the quasiparticles remaining coupled to the electromagnetic field as 
in \eqn{LBEq}. This strong difference between $\bd{v}(\bk)$ and the transport velocity $\overline{\bd{v}}(\bk)$, thus between the effective mass and the optical one, was explicitly uncovered by Grilli and Kotliar \cite{Gabi&Marco-PRL1990} in the $t-J$ model using the slave-boson formalism and including quantum fluctuations beyond the saddle-point by a large-$N$ expansion.      
\\ 

\noindent
The next question is whether these quasiparticles that do not contribute to the Drude weight can, nonetheless, be coupled to a magnetic field. In absence of electric field, we can again drop the dependence upon $\br$, thus Eq.~\eqn{LBEq} 
in frequency space becomes 
\beal
-i\omega\,\delta n_{\bk}(\omega) &=\fract{e}{c}\;\bd{v}(\bk)\wedge\bd{B}\cdot
\bd{\nabla}_\bk\,\overline{\delta n}_{\bk}(\omega) = 
-\fract{e}{c}\;\bd{B}\cdot\bd{v}(\bk)\wedge
\bd{\nabla}_\bk\,\delta n_{\bk}(\omega)\\
&\qquad
+\fract{e}{c}\;\fract{\partial f\big(\ep_*(\bk)\big)}{\partial\ep_*(\bk)}\;\bd{B}\cdot\bd{v}(\bk)\wedge
\bd{\nabla}_\bk\,\Bigg\{
\fract{2}{V}\,\sum_{\bk'}\,f_{S\,\bk\bk'}\,\delta n_{\bkp}(\omega)\Bigg\}
 \,.
\label{LBEq-1}
\eal
The frequencies $\omega$ that solve \eqn{LBEq-1} define the cyclotron resonance frequencies. Therefore, should they be finite, we would conclude that the quasiparticles do couple to $\bd{B}$.\\
Let us assume a deviation from equilibrium localised around the quasiparticle Fermi surface, which we can parametrise 
as
\bealn
\delta n_{\bk}(\omega) = - \fract{\partial f\big(\ep_*(\bk)\big)}{\partial\ep_*(\bk)}\;u(\omega,\bk)\,,
\eal
where $u(\omega,\bk)$ is defined for $\bk$ running along the Fermi surface and satisfying  
\beal
-i\omega\,u(\omega,\bk) &=
-\fract{e}{c}\;\bd{B}\cdot\bd{v}(\bk)\wedge
\bd{\nabla}_\bk\,u(\omega,\bk)\\
&\qquad
+\fract{e}{c}\;\bd{B}\cdot\bd{v}(\bk)\wedge
\bd{\nabla}_\bk\,\Bigg\{
\fract{2}{V}\,\sum_{\bk'}\,f_{S\,\bk\bk'}\,\fract{\partial f\big(\ep_*(\bkp)\big)}{\partial\ep_*(\bkp)}\;u(\omega,\bkp)\Bigg\}
 \,.
 \label{LBEq-2}
\eal
Without specifying $\ep_*(\bk)$ and $f_{S\,\bk\bkp}$ we cannot solve the eigenvalue integral equation. 
However, a formal solution can be readily obtained if we assume spherical symmetry, although this 
circumstance will never be realised in any physical Mott insulator. In this case, $\ep_*(\bk)=\ep_*(k)$, 
the Fermi surface is a $d$-dimensional sphere with radius $k_f:\, \ep_*(k_F)=0$, 
$\bd{v}(\bk) = \bk/m_*$, and we can 
expand \cite{Pines&Nozieres}
\bealn
u(\omega,\bk) &= \sum_{\ell \ell_z}\, u_{\ell \ell_z}(\omega)\,Y_{\ell \ell_z}(\Omega_\bk)\,,& 
f_{S\,\bk\bkp} &= \sum_\ell\,f_{S\,\ell}\, P_\ell\big(\theta_{\bk\bkp}\big)\,. 
\eal
in terms of spherical harmonics $Y_{\ell \ell_z}$ and Legendre polynomials $P_\ell$, 
where $\Omega_\bk$ is the angle that defines the direction of $\bk$ and $\theta_{\bk\bkp}$ 
the angle between $\bk$ and $\bkp$. Since $\bk\wedge\bd{\nabla}_\bk = i\,\bd{L}_\bk$, where $\bd{L}_\bk$ is the angular momentum 
operator in momentum space, using known properties of the spherical harmonics and 
Legendre polynomials, one readily find that \eqn{LBEq-2} becomes, taking
$\bd{B}=(0,0,B)$, 
\beal
\sum_{\ell \ell_z}\, u_{\ell m}(\omega)\,\Bigg[\omega - \fract{e}{\;m_*\,c\;}\,\bigg(1+\fract{F_\ell^S}{\;2\ell+1\;}\bigg)\,
B\,\ell_z\Bigg]\,Y_{\ell \ell_z}\big(\Omega_\bk\big) = 0\,,
\eal
where 
\bealn
F_\ell^S &= 2\rho_*\, f_{S\,\ell} = f_{S\,\ell}\,\fract{2}{V}\,\sum_\bk\,\delta\big(\ep_*(k)\big)\,,
\eal
and yields cyclotron resonance frequencies
\beal
\omega_{\ell\,\ell_z} &= \fract{e}{\;m_*\,c\;}\,\bigg(1+\fract{F^\ell_S}{\;2\ell+1\;}\bigg)\,
B\,\ell_z\,,& \ell_z=-\ell,\dots,\ell\,,
\eal
evidently defined for $\ell_z\not=0$ and corresponding to multipolar deformations of the spherical 
Fermi surface. The vanishing Drude weight in spherical symmetry amounts to 
\bealn 
\overline{\bd{v}}(\bk) = \bd{v}(\bk)\,\bigg(1+\fract{F_1^S}{3}\bigg) =0 \;\Rightarrow\;
F_1^S = -3\,, 
\eal 
which implies that only $\omega_{1\ell_z}=0$, while $\omega_{\ell\,\ell_z}$ can well be finite for $\ell>1$. This simple example shows that the quasiparticles are coupled to the magnetic field, i.e., 
they do carry an orbital magnetic moment, despite they do not contribute to charge transport. 
Even though this is not a proof, still it makes very plausible that these quasiparticles display quantum oscillations in a magnetic field, as showed more explicitly in \cite{Michele3}.

\bibliographystyle{sn-nature}
